\documentclass[]{spie}  %>>> use for US letter paper
\usepackage[dvips]{graphicx}
\usepackage{verbatim}
\usepackage{multirow}
\title{Receiver control for the Submillimeter Array} 

\author{T.R. Hunter\supit{a,b}, R.W. Wilson\supit{a}, R. Kimberk\supit{a}, P.S. Leiker\supit{a} and R. Christensen\supit{c}
\skiplinehalf
\supit{a}Harvard-Smithsonian Center for Astrophysics, Cambridge, MA, USA \\
\supit{b}National Radio Astronomy Observatory, Charlottesville, VA, USA \\
\supit{c}Smithsonian Submillimeter Array, Hilo, HI, USA
}

\authorinfo{Further author information: (Send correspondence to T.R.H.)\\T.R.H.: E-mail: thunter@nrao.edu, Telephone: 1 434 244 6836\\Address: National Radio Astronomy Observatory, 520 Edgemont Road, Charlottesville, VA 22903, USA} 
\begin{document} 
  \maketitle

\begin{abstract}

Efficient operation of a submillimeter interferometer requires remote
(preferably automated) control of mechanically tuned local
oscillators, phase-lock loops, mixers, optics, calibration vanes and
cryostats.  The present control system for these aspects of the
Submillimeter Array (SMA) will be described.  Distributed processing
forms the underlying architecture and the software is split between
hardware platforms in a leader/follower arrangement.  In each antenna
cabin, a serial network of up to ten independent 80C196
microcontroller boards attaches to the real-time PowerPC computer
(running LynxOS).  A multi-threaded, gcc-compiled leader program on
the PowerPC accepts top-level requests via remote procedure calls
(RPC), subsequently dispatches tuning commands to the relevant
follower microcontrollers, and regularly reports the system status to
optical-fiber-based reflective memory for common access by the
telescope monitor and error reporting system.  All serial
communication occurs asynchronously via encoded, variable-length
packets. The microcontrollers respond to the requested commands and
queries by accessing non-volatile, rewriteable lookup-tables (when
appropriate) and executing embedded software that operates additional
electronic devices (DACs, ADCs, etc.).  Since various receiver
hardware components require linear or rotary motion, each
microcontroller also implements a position servo via a one-millisecond
interrupt service routine which drives a DC-motor/encoder combination
that remains standard across each subsystem.

\end{abstract}


%%%%%%%%%%%%%%%%%%%%%%%%%%%%%%%%%%%%%%%%%%%%%%%%%%%%%%%%%%%%%
\section{INTRODUCTION} 
\label{sect:intro} 

The Submillimeter Array (SMA) is a collaborative project of the
Smithsonian Astrophysical Observatory (SAO) and the Academia Sinica
Institute of Astronomy \& Astrophysics of Taiwan (ASIAA).  The array
consists of eight six-meter diameter antennas with receivers operating
from 200-900 GHz and a digital correlator with 2 GHz bandwidth.  Many
design aspects of the array are analogous to the lower-frequency
millimeter interferometers completed in the last decade (Owens
Valley\cite{Padin91}, BIMA\cite{Welch96}, IRAM\cite{Guilloteau92} and
Nobeyama\cite{Morita94}).  Located on Mauna Kea, Hawaii, the primary
elements of the SMA interferometer can be reconfigured across 24 pads
which provide baselines of 8 to 500 meters.  The surface accuracy of
each dish is optimized and periodically monitored via interferometric
holography measurements \cite{Sridharan01,Sridharan02,Sridharan04}.  Further
general details on the array can be found in a summary paper
\cite{Moran98}.  Descriptions of the low-noise superconductive (SIS)
receivers\cite{Blundell98} and their associated cryostats and
optics\cite{Paine94} have been previously published. In this paper, we
describe the electrical and mechanical hardware and software that
remotely controls these instruments.  We provide a progress report on
the deployment of this system.

%%%%%%%%%%%%%%%%%%%%%%%%%%%%%%%%%%%%%%%%%%%%%%%%%%%%%%%%%%%%%
\section{COMPONENTS TO BE CONTROLLED}
\label{components}

A submillimeter interferometer presents a daunting number of hardware
components to be controlled and monitored to ensure the integrity of
the desired observations.  Since much of the equipment must be
duplicated in each individual antenna, reliable remote control of
these components is essential.  Furthermore, automated control is
desirable whenever possible in order to ease the demands on the
telescope operator.  In this section, we present a brief list of the
major receiver-related hardware components that we are controlling for
the SMA.

\subsection{Optical components}

The first concern for a tracking interferometer is the pointing of the
primary dish and subreflector.  Beyond these elements, there are
several additional optical components in the system.  The SMA is
designed to operate in (up to) two receiver bands simultaneously.  The
mechanism for selecting receivers is a combination of a
polarization-splitting wire grid and a flat mirror.  Located inside
the receiver cabin (which does not tip in elevation), these components
are rotated into specific orientations in order to illuminate the
desired pair of receivers.  Accurate positioning of their mechanical
stages is required to provide consistent and identical pointing of the
two feeds on the sky.  For observations requiring the measurement of
full Stokes parameters, a mechanism for inserting a waveplate into the
beam must be present. Finally, thermally-controlled calibration vanes
coated with millimeterwave absorbers must be moved in and out of the
receiver beam at regular intervals to provide system temperature
measurements, thereby improving the amplitude calibration of the
observations.

\begin{table}[h]
\caption{List of hardware items to be controlled and automated in the SMA
receiver system.}
\begin{center}
\begin{tabular}{|l|l|l|}
\hline
& {\bf Hardware} & {\bf Required Function(s)}\\
\hline
\multirow{5}{0.7in}{\bf Optics} 
&Polarizing wire grid & Rotate to select low-frequency band receiver \\
&Combiner mirror & Rotate to select high-frequency band receiver \\
&Quarter wave plate & Move in and out of beam\\
&Hot calibration vane & Move in and out of beam, record temperature\\
&Cold calibration vane & Move in and out of beam, record temperature\\ 
\hline
\multirow{5}{0.7in}{\bf Mixer} 
&SIS mixer selection      & Activate bias circuitry for selected pair of receivers\\ 
&SIS mixer bias V and I   & Provide set points and sense lines from bias circuitry\\ 
&SIS mixer magnetic field & Provide set points and sense lines from bias circuitry\\
&SIS total power detector & Monitor continuum levels, measure system temperature\\
&FET switch               & Select IF signals to output from cryostat\\
\hline
\multirow{5}{0.7in}{\bf LO}
&Gunn oscillator      & Tune two actuators to acquire and maintain phase-lock \\
&Harmonic mixer       & Adjust YIG reference oscillator power for best mixing\\
&Waveguide attenuator & Adjust actuator for optimal receiver sensitivity\\
&Frequency multiplier & Tune two actuators for optimal LO power\\
&Martin-Puplett diplexers & Adjust mirror spacing for maximum LO injection\\
&Digital PLL          & Enable/disable Gunn, set locking sideband, loop gain,\\
&                     & monitor phase-lock status\\ 
\hline
%\multicolumn{2}{c}{}\\
%\hline
\multirow{5}{0.7in}{\bf Cryostat}
&Vacuum       & Monitor pressure with Pirani and cold cathode devices\\
&             & Start/stop turbopump, monitor the RPM\\
&Refrigerator & Monitor refrigerator stages and receiver cold plates\\
&             & Adjust J-T needle valve to achieve 4 Kelvin\\
\hline

\end{tabular}
\end{center}
\end{table}

\subsection{Receiver components}

The frequency coverage of the SMA receivers is divided into six bands
from 200-920 GHz.  The first generation of receivers now on the
telescope include the 230, 345 and 690 bands.  Each receiver employs a
single niobium SIS mixer.  Each mixer requires a junction bias voltage
and a magnetic field bias voltage to be present and tunable.  In
addition, a local oscillator (LO) signal must be combined with the sky
signal to make the receiver sensitive to radiation.  Each LO contains
a chain of waveguide devices (Gunn oscillator, attenuator and
multiplier) which require mechanically-tuned actuators to achieve the
desired operating frequency.  A phase-lock loop (PLL)
circuit\cite{Gardner79} maintains the phase and frequency of the LO
signal in agreement with the main reference generator (MRG)
distributed to each antenna via fiber optics.  Detailed laboratory
calibration of the individual LO chain devices is an essential step
for reliable tuning and operation of the receivers.  Feedback from the
receiver continuum (total power) detector is essential to measure the
receiver sensitivity (via the Y-factor method) and thereby automate
the tuning of all these items.

\subsection{Cryostat components}

Since the SIS receivers must be operated continuously at 4K, they are
housed in an evacuated cryostat cooled by a three-stage mechanical
refrigerator utilitizing the Joule-Thomson (J-T) effect.  Full control
of the cryostat temperature and pressure requires a turbopump, vacuum
gauge, a suite of diode thermometers, and a J-T needle valve.  In
addition, a cold FET switch must be controlled to select the
appropriate pair of receiver intermediate frequency (IF) signals to be
output from the cryostat.

\section{HARDWARE ELECTRONICS}

\subsection{Design goals} 

All of the items discussed in section~\ref{components} reside inside
the antenna receiver cabin. In principal, all of the functions
described so far could be accomplished by a single computer.  However,
this arrangement would require a large number of (potentially) long
wires containing sensitive signals prone to noise pickup.  Instead, we
have chosen the principal of {\it distributed processing}, in which
the system logic is spread between a larger number of simpler
computers located physically close to the item for which they are
responsible.  There are many advantages of distributed processing,
primarily the greater modularity in each device.  Both the hardware
and software component of a device can be quickly and simultaneously
replaced with a spare, which is an important quality at a remote site
with many copies of the same hardware (i.e. an interferometer).  Also,
hardware specific information (such as calibration tables) can be
stored at a low-level rather than being edited and propagated down
from a large, high-level storage area.  The main disadvantage of
distributed processing is the greater number of computers which must
communicate in a sensible and cooperative way.  Also, global software
updates must be propagated to many more devices.  In a sense,
distributed processing moves the complexity from the wiring to the
software, which is more feasible to debug and augment from a remote
location.

\subsection{Microcontroller circuit boards} 
\label{micros}

The design of the principal set of microcontroller circuit boards
began in 1995, and production and testing has been ongoing since then.
The set of boards is divided into three categories: LO control board,
mixer control board and optics control board.  Each microcontroller
board includes an Intel 80C196KC 16-bit microcontroller running at
20MHz with an 8-bit bus, along with additional hardware specific to
the board's function.  This microcontroller is a versatile device that
can address up to 64 kilobytes of memory using a von Neumann
architecture.  In our case, the boards each contain a 32 kilobyte
EPROM and two banks of 8 kilobyte nonvolatile static random access
memory (NVSRAM). Each antenna is outfitted with a single mixer control board
and optics control board, and up to 8 LO control boards.  Using multi-layer
surface-mount technology, the boards measure only a few inches on a
side.  All of them reside on a single RS485 serial bus managed by a
multi-threaded leader program running on the antenna computer--a
diskless PowerPC controller (Motorola MVME2700) with a commercial
real-time Unix operating system (LynxOS).  A related microcontroller
board, called the servo control board, employs the same 80C196KC
device to run the velocity servo loop and system status monitor of the
main antenna drive motors.  The SMA servo control board provides a
good example of distributed control.  Additional details have been
presented elsewhere \cite{Hunter01,Hunter13} and will not be further discussed
here.

\subsubsection{Local oscillator control board}

The LO control board contains motor driver circuits for up to eight DC
motors.  The eight channels are allotted as follows: two for the Gunn
oscillator, one for the attenuator, up to four for the multiplier(s)
and one for a Martin-Puplett diplexer (to provide optimal LO injection
in the high frequency bands).  Each motor has a rotary encoder read by
a quadrature decoder/counter interface on the LO control board.
Associated with each LO control board is a digital PLL (DPLL) circuit
and a thermoelectric controller (TEC) to power and control the Gunn
oscillator.

\subsubsection{Digital phase-lock loop}
\label{dpll}

The DPLL is based on a high-gain second-order loop with an active
filter.  The IF signal from the harmonic mixer is first amplified by
60dB (across a bandwidth of 0-500MHz) and then passed to an Analog
Devices AD96687 Ultrafast ECL Dual Comparator which converts the
sinewave to a squarewave.  Zero crossing detection is used in the
comparator in order to avoid an automatic gain control (AGC) circuit
in the IF amplifier.  By biasing the comparator above the peak value
of the IF signal, the microcontroller can disable the loop.  The
digitized signal output from the comparator enters the Analog Devices
AD9901 ultrahigh speed phase/frequency discriminator.  An AD817 high
speed, low-power wide supply range amplifier provides the loop
integrator.  We implement a loop gain control by controlling the
current in the AD9901 via the 8-bit DAC in the microcontroller.  A
second AD817 monitors the Gunn bias voltage and, along with a diode
pair, keeps it within a diode drop ($\pm 0.6$~V) of the target bias
voltage set manually via a potentiometer.  The DPLL provides a high
loop bandwidth (a few MHz) in order to reacquire lock after each phase
switch in the 109MHz reference signal due to the interferometer Walsh
cycle\cite{Thompson01} generated by the direct digital synthesizer
(DDS).  The eight-channel 10-bit ADC in the LO microcontroller is used
to monitor the Gunn bias, a temperature sensor and the power supply
voltages.  The high speed output of the microcontroller is used to
control the on/off state of the Gunn oscillator and DPLL, the sideband
of the DPLL, and a heater resistor to keep the diplexer section of the
DPLL warm when the Gunn oscillator is not in use (thereby avoiding
initial phase drifts at startup).

\subsubsection{Optics control board}

Like the LO boards, the optics control board has motor driver circuits
for up to eight motors, two of which are higher current capacity
drivers (for the grid and combiner mirror).  The board also contains
eight digital inputs attached to the optical breaks which provide home
positions for the various rotating and translating stages.  The
eight-channel 10-bit ADC inside the microcontroller is used to monitor
the calibration vane temperatures.

\subsubsection{Mixer control board}

The mixer control board contains four 12-bit DACs to provide the mixer
bias and magnetic field command voltages to the active pair of
receiver bias circuits, and a ten-way multiplexed 12-bit ADC to
monitor the mixer bias, mixer current, magnetic field, receiver total
power and other voltages.  A set of digital outputs is used to select
the two active receivers, which includes activating the appropriate
quadrant of the mixer bias circuitry and setting the cryogenic FET IF
switch.  The next revision of the board will also contain a motor
driver for the cryostat J-T valve.

\subsection{Other devices}

In contrast to the receiver and optics components, the cryostat
instrumentation (turbo pump, vacuum gauges and thermometers) is
controlled by commercial RS-232 interface units sold by the individual
manufacturers (Varian, Balzers and Lakeshore).  Each of these discrete
units is managed by a separate monitor program running on the antenna
computer.

\begin{comment}
The optical CCD guidescope for the SMA contains an
Electrim-104N CCD.  We have attached the interface card to an Intel
486 PC in PC-104 format running Red Hat Linux 6.2 and the camera
device driver from the manufactuer.  Like the antenna computers, this
machine boots over the network and mounts a disk drive on which it
records images.  A separate device driver controls the camera
single-axis focus stage, along with the telescope tube temperature and
humidity sensors.
\end{comment}

\section{ONLINE SOFTWARE}

\begin{figure}[hb]
   \begin{center}
   \begin{tabular}{c}
   \includegraphics[width=5.50in]{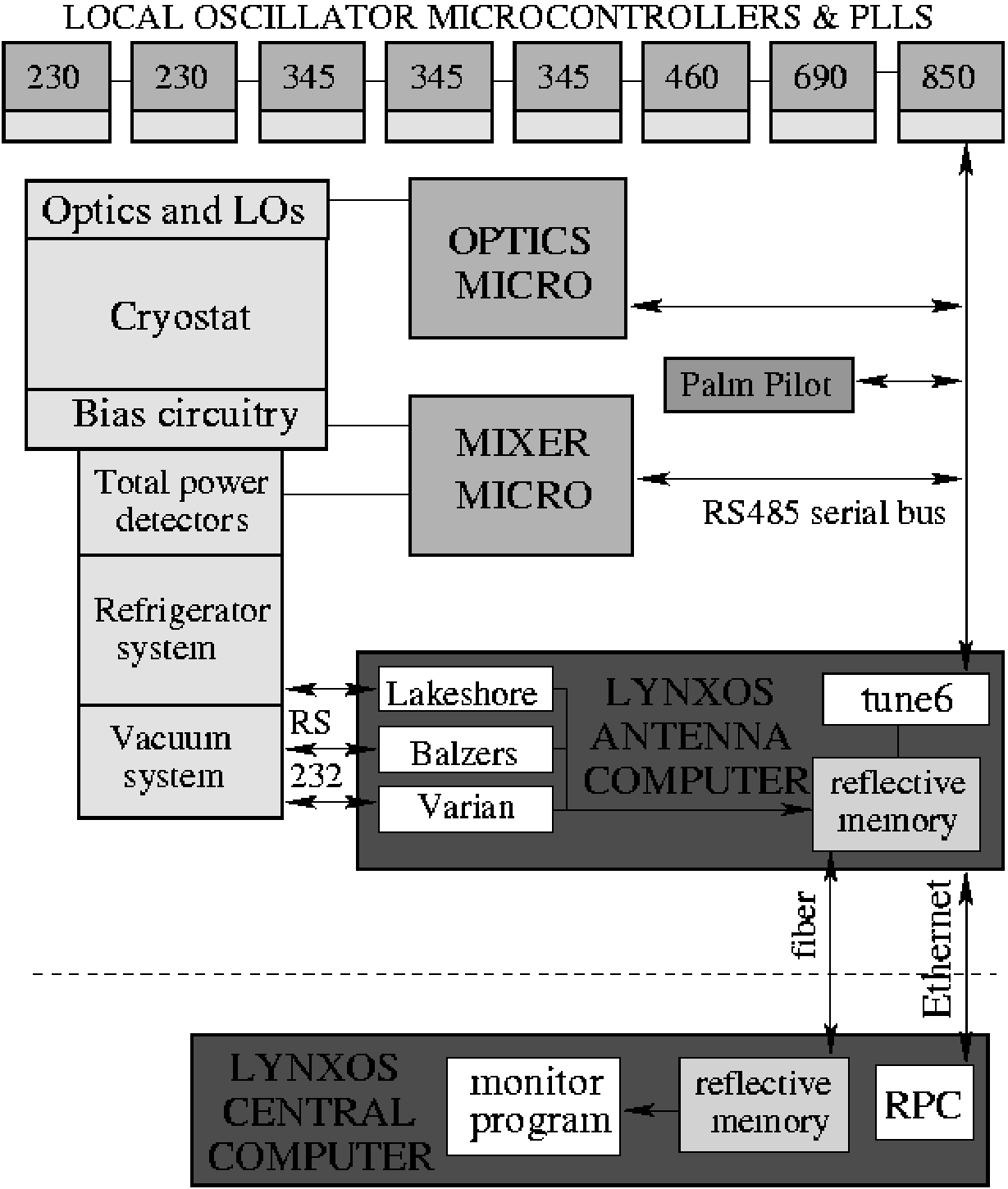}
   \end{tabular}
   \end{center}
   \caption[example] 
   { \label{fig:layout} 

Layout of the SMA receiver control system.  Commands come from the
central computer via Ethernet, either in RPC format or from an
interactive console option.  To fulfill these commands, the leader
server program ``tune6'' dispatches appropriate sub-commands to the
microcontroller boards via a common RS485 bus, while constantly
logging the system status to reflective memory.

}
   \end{figure}

The online software structure of the SMA contains many levels of
monitor and control functions.  A central element in the organization
is a fiber-linked hardware reflective memory (RM) system (VME
Microsystems International Corporation).  Critical monitor points are
regularly logged to predefined locations in RM by various
low-level programs running in the antenna computers, correlater crate
computers, etc.  The RM variables are read by higher-level status
display programs (both text-based and graphical) and the error
reporting system.  For the most part, control commands are sent from
the central computer (hal9000) to individual antenna computers via the
remote procedure call (RPC) protocol\cite{Bloomer92}, in some cases
bundled into Perl scripts.  In the case of the receiver system, these 
commands are served by two layers of programs: a leader server program
on the antenna computer, and the set of follower programs running on the
microcontrollers that it manages (see Figure~\ref{fig:layout}).  All
programming is done in C, using a cross-compiler running on Solaris
(or Linux) for the PPC-LynxOS code and the Tasking Inc. compiler
running on Solaris for the Intel microcontroller code.

\subsection{Antenna computer leader program} 

A leader server program (called tune6) runs on each antenna computer
and talks to each of the microcontroller boards described in
section~\ref{micros}.  Tune6 has matured into a powerful,
multi-threaded program.  At startup time, it probes the RS485 serial
bus to discover which boards are present, and thereby what capabilities
are available to the user.  In addition to the RPC server thread, it
has an optional interactive command-line thread that is used primarily
in laboratory situations for hardware calibration functions, but is
also quite useful in debugging problems on-site.  A separate thread
monitors critical values on each of the microcontroller boards into
reflective memory at several Hertz.  Access to the serial line amongst
the threads is regulated by a mutex.  In addition to the serial line,
tune6 can talk to a dozen different GPIB devices (spectrum analyzer,
synthesizer, multimeter, function generator, power meter, etc.).
Tune6 can trigger resets of the microcontroller boards via a digital
I/O card.  It also controls the output power of the YIG oscillators
and the RF switches that direct the YIG signal to the various
receivers.  It can acquire and simultaneously display current
vs. voltage (I/V) and power vs. voltage (P/V) sweeps to any IP address
via Tcl/Tk scripts\cite{Osterhout94} and the wish shell, without the
need for installing either Xwindows or Motif on the antenna computer,
which was an undesirable prerequisite.

An alternate leader program running on a Palm pilot can also control
the bus in a ``local mode'', similar to the one on the antenna motor
drive system.  Because it resides on the same serial bus, it can
eavesdrop on all the packets and provide a visual display of the
system status even when the the antenna computer is in control.  We
use the gnu prc-tools cross-compiler on a Linux machine to generate
the Palm pilot executable\cite{Rhodes92}.

\subsection{Microcontroller follower programs}

The goal for the microcontroller software is to implement hardware
specific functions that can be called in succession when needed by the
leader program tune6.  Most of the embedded microcontroller code
resides in the EPROM.  The board type identifier is stored here along
with the communication subroutines, which are a common object module
to all boards.  Since each board sees all packets on the RS485 bus,
the packets contain the source and destination addresses, thus they
are ignored by all boards except the target.  (See
appendix~\ref{packetprotocol} for a detailed description of the packet
protocol.)  Each packet is encoded with a cyclic redundancy check
(CRC) so that transmission errors can be identified.  In practice, the
low bandwidth used (38.4kbaud) makes errors very rare.  The
``operating system'' running on the microcontroller boards is a simple
command dispatcher, with essentially zero boot time.  There are no
priorities assigned to jobs--they are completed in the order received.
A board which has received a command essentially has ``control'' of
the RS485 bus until it has completed its command and issued an
appropriate response to the caller, which is typically tune6, but can
be another board previously commanded by tune6.

The lower section of NVSRAM is used for data tables. All hardware
specific data, such as frequency tuning lookup tables and rotary stage
positions, are stored here, but can be easily reloaded or modified by
tune6 commands.  In the case of the LO boards, the board sub-type
(i.e. the frequency band) is stored here.  This section of memory is
automatically stored at power-down, and restored at power-up.  The
higher section of NVSRAM is used for additional embedded program code.
New software features are generally programmed into this area first
for debugging prior to being moved to the permanent EPROM.

\subsection{DC motor servo}

The LO and optics microcontroller board software each contain an
identical interrupt-driven servo.  The hardware allows two of the
eight motors can be active at a time, though for software simplicity
we run only one of them at a time.  When a new position is commanded,
the calling function on the microcontroller waits until the position
has been acquired.  We use the microcontroller software timer to
trigger the interrupt service routine every millisecond.  The encoders
are read and the velocities during the past millisecond are
calculated.  The present position error is computed and adjusted by
velocity feedback.  If the position error changes sign, the motor
direction is reversed.  If the position error is zero, the motor
current is stopped.  As soon as the position error remains below 2
encoder counts for 20 consecutive cycles, the calling function
proceeds.

\section{CALIBRATION SOFTWARE}

A large portion of the effort in this project has gone into developing
automated calibration software for the various components of the
system.  Detailed laboratory calibration of millimeter oscillator
hardware is particularly crucial in order to achieve reliable
operation at the telescope\cite{Thatte95}.

\subsection{LO boards}

\begin{figure}[hb]
   \begin{center}
   \begin{tabular}{c}
   \includegraphics[width=6.0in]{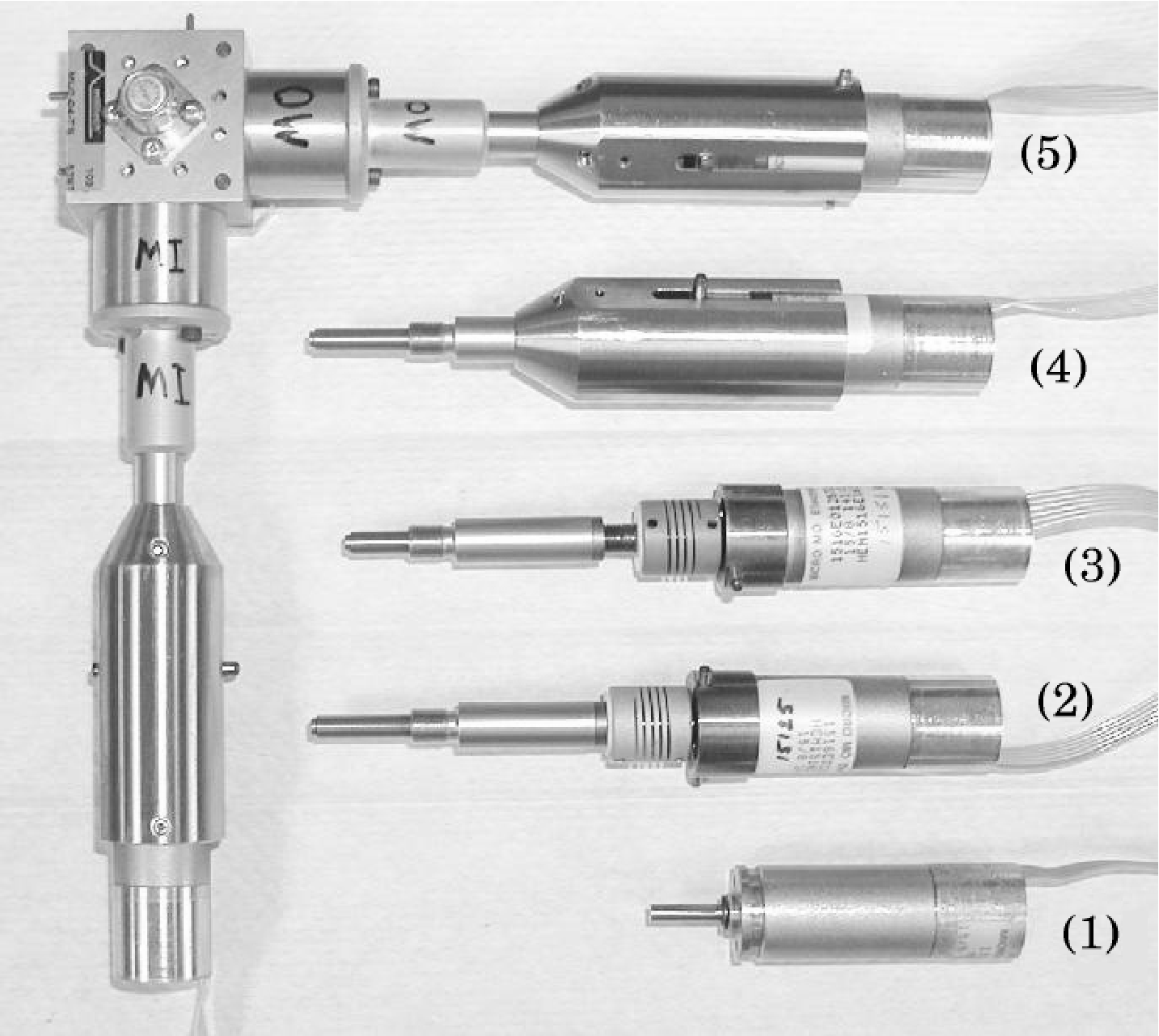}
   \end{tabular}
   \end{center}
   \caption[example] 
   { \label{fig:actuator} 
The linear actuator used to tune the spring-loaded waveguide plungers 
in the Gunn oscillator, attenuator and multiplier consists of a
rotary DC motor with an encoder and a flexible coupling to a leadscrew.
(1) DC motor and encoder unit,
(2) Actuator in the fully-extended state, with the bearing against
the teflon washer hardstop,
(3) Actuator in a partially-retracted state,
(4) Actuator assembled with shell,
(5) Two actuators on an assembled multiplier.
}
   \end{figure}

In our system, the microwave LO chain hardware requires the largest
amount of calibration effort.  On the Gunn oscillator, attenuator and
multiplier, the manual micrometers are first removed and replaced by
linear actuators.  The Klinger actuators (Figure~\ref{fig:actuator})
are comprised of a MicroMo rotary motor with an encoder and a
leadscrew attached through a flexible coupling.  Appropriate width
washers are machined until the throw of the actuator matches the
micrometer.  The fully extended position is used as the reference hard
stop for the encoder. A teflon washer is included to prevent the
actuator from jamming at the hard stop.  The applied motor torque
heading toward the stop is reduced using a zener diode on the voltage
drive line.  With 64 pulses per revolution, a 141:1 gear ratio and a
screw pitch of 2 threads per millimeter, the resolution of the encoder
corresponds to a linear motion of .05 micron.  When the servo acquires
a specified rotary encoder position from opposite directions, the
achieved linear position exhibits a consistent difference of about 5
microns. In repeated motions, the relative position change is better
than 2 microns.

\subsubsection{Gunn oscillator calibration}

\begin{figure}[hb]
   \begin{center}
   \begin{tabular}{c}
   \includegraphics[width=6.5in]{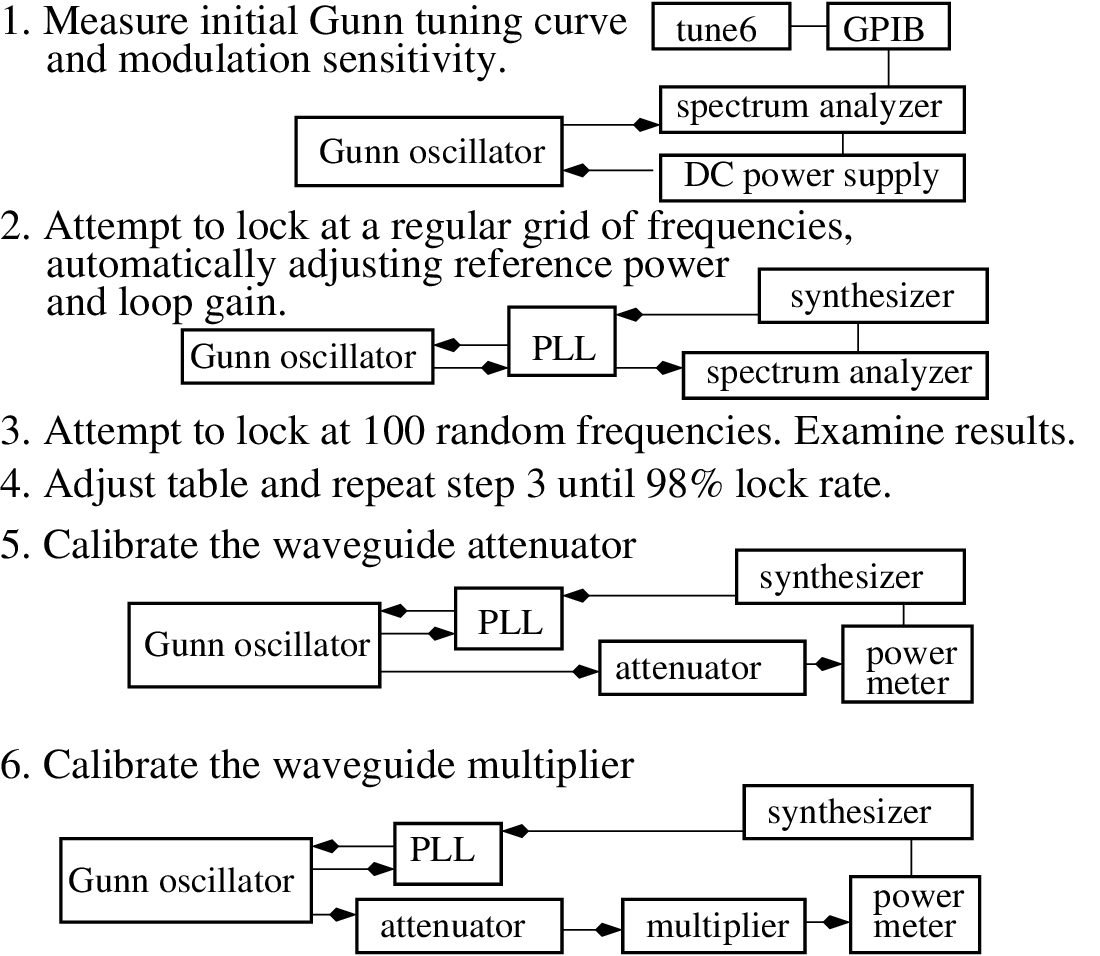}
   \end{tabular}
   \end{center}
   \caption[example] 
   { \label{fig:gunncalib} 
Major calibration steps for a motorized submillimeter local oscillator chain.
}
   \end{figure}

The frequency output of a Gunn oscillator depends on both the applied
bias voltage and the length of a mechanical tuning
cavity\cite{Carlstrom85}.  The calibration sequence of the Gunn
oscillator (see Figure~\ref{fig:gunncalib}) begins with a fixed bias
voltage applied.  The tuner and power backshort actuators are scanned,
with the signal identification function running on a spectrum analyzer
equipped with an external mixer.  An initial tuning table of tuner
motor position vs. frequency is thus generated.  If a programmable
power supply is available, the Gunn modulation sensitivity (typically
$\sim 100-400$MHz/volt) is also measured during this procedure at each
frequency by applying a small change in voltage and observing the 
change in frequency.  It is important to identify any frequencies
where the sensitivity approaches zero as they will be difficult (or
impossible) to phase lock.  The Gunn tuning table can be typically fit
quite accurately with a monotonic segment of a fifth-order polynomial.
The series of backshort settings that generate a local maximum in 
output power is analyzed by an algorithm which tries to select a
sensible, continuous mode.  Frequencies at which insufficent power is
available to drive the multiplier are identified as holes in the
tuning table.  The tuning table is then downloaded to the
microcontroller and a automatic tuning test of 100 random frequencies
is initiated, using the phase-lock search algorithm.

\subsubsection{Phase-lock search algorithm}

When a frequency command is given to the tune6 server, it is passed to
the appropriate LO control board which examines its local lookup table
and computes interpolated positions for the Gunn and multiplier
backshorts.  Once the actuators reach these positions, the DPLL
voltages are examined.  If there is no phase lock, a search mode
begins.  The Gunn tuner is scanned across the expected position.  By
the nature of the DPLL design (section~\ref{dpll}), phase lock is
indicated whenever the Gunn bias voltage lies between the two voltage
limits established by the diode pair.  However, if the IF amplifier is
being overdriven, it is possible to ``lock'' on a harmonic of the IF.
To reject these false ``locks'', two LC circuits are included which
provide a simple yet powerful analysis of the IF spectrum to the
tuning software.  A portion of the IF signal is tapped and sent
through a notch filter tuned to 109 MHz and on to a Schottky diode
pair.  A similar circuit implements a bandpass filter tuned to the
same frequency.  The Gunn bias and the voltage outputs of both filter
circuits are digitized by the microcontroller board and used in the
automated phaselock algorithm.  Proper lock is indicated by a large
ratio of bandpass to notch power combined with a Gunn bias that is
less than a diode drop from the target.  Once the Gunn is locked, the
bias voltage may slowly drift as the PLL compensates for any residual
thermally-induced changes in the tuner cavity length.  Whenever the
voltage drift exceeds half a diode drop from the target value, the
mechanical tuner is automatically adjusted to recenter the bias.  This
feature maintains lock during astronomical observations.

\subsubsection{Tuning table generation}

During the calibration tests of a new Gunn, the initial lock search
takes place using default values for the reference power and loop
gain.  If the search fails, these values are adjusted automatically
and the search begins again.  At all frequencies where lock is
ultimately achieved, the power and gain settings are stored.  The
signal to noise of the IF trace is also recorded as a function of the
reference signal power and the loop gain.  With this information, a
better tuning table can be constructed in which the reference power
and gain vary with frequency.  The varying reference power requirement
is due to the variation of the harmonic mixer efficiency as a function
of both frequency and applied power.  The varying gain requirement
results from the variation of the modulation sensitivity of the Gunn
with frequency.  After a few appropriate adjustments are made to the
tuning table, the test is repeated until 99 percent success rate is
achieved.  There are tune6 commands to convert an ASCII-delimited
tuning table to Intel hex format suitable for download to the
microcontrollers.  Commands to upload and plot the existing tables
also exist.  Single line changes to the table entries can also be
applied directly to the on-line microcontroller memory.  The frequency
increment between entries need not be constant throughout the table,
though we find that 1 GHz is usually sufficient.

\subsubsection{Attenuator calibration}

The attenuator calibration is quite simple as it requires only a Gunn
oscillator and a W-band power meter.  The attenuation is recorded in a
single scan of the attenuator motor and the resulting table is
resampled into the minimum number of points (typically a dozen) from
which a linear interpolation will yield less than a specified error
(e.g. 0.5 dB) in the achieved value.  The resampled table is stored in
the microcontroller NVSRAM immediately following the Gunn tuning
table.  All of the tables have variable lengths with computed pointers
to the various component sections.

\subsubsection{Multiplier calibration}

\begin{figure}[hbt] 
   \begin{center}
   \begin{tabular}{c}
   \includegraphics[width=6.0in]{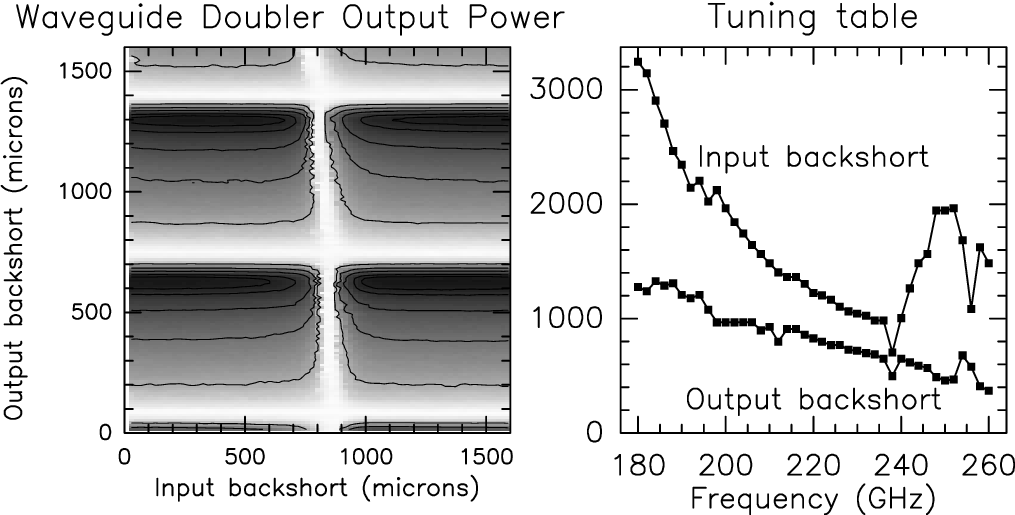}
   \end{tabular}
   \end{center}
   \caption[example] 
   { \label{fig:multipliertable} 
Left panel: Multiplier output as a function of backshort position at a
single frequency (246 GHz).  Contour levels: 20,40,60,80,90\%. Right panel: Optimum backshort positions as a function of frequency.
}
   \end{figure}

The frequency multipliers are operating with passive, self-bias to
protect them from static discharge, and to simplify the tuning.  The
calibration sequence of a multiplier requires either a high-frequency
total power detector or an operating SIS receiver with a total power
detector.  First, the Gunn is tuned to the highest available
frequency, and the multiplier backshorts are individually scanned
through their full range.  When the peak power is identified, the Gunn
frequency is lowered to the next setting and a local maximum is found
on both multiplier backshorts.  The results of this procedure form the
lookup table for the multiplier (see
Figure~\ref{fig:multipliertable}).  This table is stored in memory
after the attenuator table.  In practice, the final optimization of
multiplier output power is done with feedback from the receiver total
power detector.  

\subsection{Optics control board}

Some of the optical components in the SMA require precise positioning
while others need only coarse motion.  The polarization-splitting
wire-grid and the combiner mirror are two elements that are critical
to the receiver alignment.  In each case, the alignment is carefully
measured in the lab and a mechanical detent is placed at each receiver
port so that the rotating stage can be locked into place manually.
Once the motor drive gear has been engaged, the same detent is used by
the automatic drive system.  Like the LO chain motors, the grid motor
has an encoder attached.  In this case, one encoder tick corresponds
to a 9.5 arcsecond rotation of the grid stage.  For the calibration
sequence, the stage is driven in small steps until the detent is
engaged.  The encoder position is noted at the position of each detent
and stored in memory.  When the optics control board receives a command to
select a new receiver, the servo drives the grid motor until the
stored encoder position is achieved.  As soon as the motor is
depowered, if the stage is within a millimeter of the detent center,
the mechanical force of the detent pulls the stage into position.
Once a good encoder position is found for each detent, this limited
feedback system works reliably.  A single optical break, triggered by
a fixed mechanical flag, is sufficient to provide a known absolute
position for the encoder.  Since the stage is rotated in only one
direction, this break is triggered once per rotation, therefore no
encoder drift can build up, and the system will automatically
restablish the rotary coordinate system after any manual moves.

The other item presently under control of the optics control board is the
first generation of the SMA receiver calibration system.  It consists
of a single, ambient-temperature vane nine square inches in size that
is moved in and out of the receiver beam by a rotary motor with a slip
clutch.  A hardstop defines the ``in'' and ``out'' positions.  Due to
the non-precision nature of the motion, a potentiometer on the motor
(rather than an encoder) is sufficient to determine when one of the
two positions is reached at which point the motor is turned off.  A
self-calibration sequence can quickly remeasure the two positions if
the unit senses any mechanical drift.

\subsection{Mixer control board}

The DAC and ADC zero offsets for the mixer bias control and feedback
lines can be measured and stored in the microcontroller memory.  This
feature ensures symmetry about the origin of the I/V curves.  The I/V
curves are not stored on the mixer control board, but rather are sent to a
hard drive in the control building that is mounted by the antenna
computer.  A modelling subroutine fits the various regions of the I/V
curve to determine the junction temperature, gap voltage and contact
resistance.  The noise temperature of the first stage amplifier is
determined from the P/V curve.  Commands exist for the mixer control board to
optimize the receiver total power output as a function of the mixer
bias and LO attenuation.  The magnetic field bias can also be scanned
to minimize the area under a specified region of the P/V curve to
maximize the receiver stability.  The mixer control board NVSRAM contains a
table to record the optimal mixer bias, magnetic field and LO
attenuation settings versus sky frequency.  These can be measured and
recorded in situ at the telescope as new frequencies are observed for
the first time.

\section{PRESENT DEPLOYMENT STATUS}

As of August 2002, seven SMA antennas are present on Mauna Kea and
four of them are undergoing routine test usage on the sky.  The optics
control board is present in all four of these antennas and is
controlling the ambient temperature calibration vane and the wire grid
rotation stage.  The mixer control board is present in all four
antennas and is fully functional.  The receiver can be tuned
interactively by computer commands and characteristic I/V curves are
routinely obtained and displayed remotely.  There are three LO control
boards and DPLLs in each antenna, one each for the 230, 345 and 690
GHz receivers.  Only the 230 GHz LO chains are motorized at this time.
The other bands must still be tuned manually.  Most of the development
work is presently aimed at motorizing and fully automating the 345 GHz
LO chains.  The Palm pilot software is partly developed and is in use
only in the laboratory.
% and occasionally in one antenna.  
Further work
on this software will resume once more of the higher priority LO chain
hardware work is completed.

\section{FUTURE REVISIONS}

The main circuit board revision to be undertaken is an upgrade of the
mixer control board.  New functionality will be added to control the
cryostat J-T valve.  Also, a total power stabilization servo will be
implemented by reading the receiver temperature and adjusting the HEMT
bias accordingly.  This feature will require higher precision ADCs and
DACs than are presently on the board.  An LED status board will be
added to provide a simple description of the present state of the
receivers to personnel entering the antenna cabin.  Other revisions
may be necessary to the connector portion of the LO control board to
accomodate the geometry of the higher frequency LO chains.  Software
control of the Martin-Puplett diplexer on the 690 GHz LO chain also
remains to be implemented.  The waveplate mechanism and the next
generation of calibration vanes (to provide a more accurate
two-temperature calibration) are still under hardware design.
Finally, some modifications to the tune6 software are being undertaken
to streamline the system for dual-frequency operation in the near
future.  The ultimate goal for this software is to have all of the
various receiver tuning commands to be triggered by a single command,
the response to which will be the receiver temperature achieved.

\acknowledgments     %>>>> equivalent to \section*{ACKNOWLEDGMENTS}       
 
We thank Ray Blundell, Hugh Gibson, Ferdinand Patt, Edward Tong and
other members of the SMA staff for useful suggestions on the system as
it has been developed and deployed on Mauna Kea.

%%%%% References %%%%%

\bibliography{article}   %>>>> bibliography data in report.bib
\bibliographystyle{spiebib}   %>>>> makes bibtex use spiebib.bst

\appendix    %>>>> this command starts appendixes
\section{SERIAL PACKET PROTOCOL}
\label{packetprotocol}

The antenna computer communicates with the series of microcontroller
boards via variable-length serial packets transmitted over an RS485
bus.  Here we give a description of the packet protocol.  The first two
bytes in each packet constitute the packet header.  The first byte of
the packet header presents a unique pattern: it is the only byte in
which the two most significant bits (MSB) are high.  The next two bits
are unused, while the four least significant bits (LSB) define the
target address.  The second byte of the packet header contains the
source address in the four LSB.  The third MSB is also set high in
order to distinguish all bytes sent from board address 10 from the
terminating character.  The other three bits are unused.  The packet
type byte plus (up to) 32 bytes of content (aside from the header,
terminator and extra encoding bytes) may be transmitted in a single
packet.  This provides space for eight long integers (in particular,
eight motor encoder values).  A global enumeration of packet types is
shared by all boards, although not all boards are programmed to
respond to all packet types.  
Following the (encoded) packet type and content bytes is the cyclic
redundancy check (CRC) word (a two-byte short integer).  Once the
packet has been assembled, the packet data are encoded, beginning with
the third byte and continuing through the CRC word.  The CRC is
performed over the entire packet, including the first two bytes which
are not encoded.  The first two bytes of the packet are not encoded so
that each board on the bus can immediately discern to whom the packet
is destined without having to decode it.  The encoding proceeds as
follows: for each byte, the sign bit is removed (and stored), and 0x20
is added to the value.  After six such bytes, a byte containing the
sign bits is constructed and placed prior to the six content bytes in
the data stream.  The value of this ``sign byte'' is 0x40 plus the
sign bits which reside in the six LSB.  The sign bit of the first byte
resides in the third MSB.  The sign bit of the second byte resides in
the fourth MSB, etc.  This cycle repeats for the rest of the data,
including the CRC word.  The final byte of the packet is the
terminator character (a ``newline'' in C code = 0x0a = 10 = 00001010).
In general, the response to a packet has the same packet type as the
initiating packet.  The contents vary as a function of the packet
type.
As far as addressing goes, three address dipswitches are located on
each microcontroller board.  The LO microcontrollers use the address
range 0-7.  Using a software implied ``8''-bit, the mixer and optics
control boards can use any two unique addresses from 8-13.  The antenna
computer program (tune6) uses address 15 and the Palm pilot controller
program uses address 14.

\end{document}